# Generalized-impedance and Stability Criterion for Grid-connected Converters

Huanhai Xin, Ziheng Li, Wei Dong, Leiqi Zhang, Zhen Wang and Jian Zhao

*Abstract*—The output impedance matrix of a grid-connected converter plays an important role in analyzing system stability. Due to the dynamics of the DC-link control and the phase locked loop (PLL), the output impedance matrices of the converter and grid are difficult to be diagonally decoupled simultaneously, neither in the dq domain nor in the phase domain. It weakens the effectiveness of impedance-based stability criterion (ISC) in system oscillation analysis. To this end, this paper innovatively proposes the generalized-impedance based stability criterion (GISC) to reduce the dimension of the transfer function matrix and simplify system small-signal stability analysis. Firstly, the impedances of the converter and the grid in polar coordinates are formulated, and the concept of generalized-impedance of the converter and the grid is put forward. Secondly, through strict mathematical derivation, the equation that implies the dynamic interaction between the converter and the grid is then extracted from the characteristic equation of the grid-connected converter system. Using the proposed method, the small-signal instability of system can be interpreted as the resonance of the generalized-impedances of the converter and the grid. Besides, the GISC is equivalent to ISC when the dynamics of the outer-loop control and PLL are not considered. Finally, the effectiveness of the proposed method is further verified using the MATLAB based digital simulation and RT-LAB based hardware-in-the-loop (HIL) simulation.

## I. Introduction

WITH the rapid growth of the renewable sources and the ongoing demand for flexibility and controllability of the power system, voltage source converters (VSCs) are increasingly used in electric power grids for the connection of wind power generations, photovoltaic power generations, and high-voltage dc (HVDC) transmissions [1-3].

The massive integration of converters has changed the dynamic characteristics of the power grid. High penetration of converter connected resources enlarges the equivalent grid impedance and makes the AC grid relatively weak. New oscillation issues arise and poses challenge to power system stability due to the complex interaction between the power electronic equipment and traditional AC grid, which becomes even more critical in a weak AC grid [4-6].

In recent years, a number of oscillation events caused by power electronic equipment occur around the world. For example, oscillations occurred in series compensated wind farm in Texas USA and the subsynchronous torsional interaction between the generators and the HVDC system happened in Northeast China Grid [7-8]. Due to the complex control system of the VSC, the mechanism of the VSC-involved oscillation is more complex and the corresponding stability is more difficult to be analyzed compared with the oscillation problems in a traditional power system [9].

The oscillation analysis of a grid-connected converter is usually studied with small-signal analysis, which can be classified into two categories, i.e., the state space based eigenvalue analysis [10-11] and the impedance-based analysis in the frequency domain [12-20]. For eigenvalue analysis, despite of its wide application in conventional power system stability, the detailed state space model and its parameters are essential, but in the real power system with large-scale renewable integrated the model inaccuracy arising from uncertainty will lead to analysis difficulties. On the other hand, in impedance-based analysis the converter and the grid can be regarded as two independent systems, whose port characteristic is measurable and thus can be used for quantitative stability analysis. Hence, it has drawn great attention and been widely applied for power system stability analysis in the past decades.

Using the impedance-based methods, the grid-connected VSC and the AC grid are both modeled as an output impedance connected to a voltage/current source. For a three-phase system, the impedance of the VSC or the grid can be mathematically represented as a matrix that describes the relationship between the voltage vector and the current vector. Currently, most impedance-based methods for grid-tied VSCs can be classified into two categories according to the reference frames to formulate impedance matrices: the synchronous frame (dq frame) based methods [13-16] and the stationary frame (abc frame) based methods [17-19].

In dq frame, there usually exist coupling terms (off-diagonal terms) in the impedance matrices of the VSC and the grid, which means that the system is MIMO. The generalized Nyquist stability criterion (GNSC) is usually used for stability performance analysis in these methods [16]. In abc frame, coupling terms also exist between the positive and the negative sequence impedances of the system. Although the two sequence impedances of a symmetrical grid can be completely decoupled (i.e., the impedance matrix is diagonal), the impedance matrix of the VSC usually has coupling terms [18]. As an approximation, the coupling terms are generally omitted to build up an equivalent SISO system so that the impedance-based stability criterion can be applicable [17, 19].

The authors are with the College of Electrical Engineering, Zhejiang University, Hangzhou 310027, China. Email: xinhh@zju.edu.cn.

However, neglecting the coupling terms may lead to inaccurate analysis results in a mirror frequency coupled system [18, 21].

This paper presents a generalized-impedance based stability criterion (GISC) for small-signal stability analysis of three-phase grid-connected VSCs. We firstly formulate the mathematical impedance model of VSCs and grid in dq frame, and then defines the concept of generalized-impedance, based on which, the stability criterion is finally proposed. In this regard, the VSCs and grid system is simplified as a SISO system with the coupling terms retained and its complex dynamic interaction mechanism can be revealed based on the oscillation analysis of generalized-impedance. Research analysis indicates that the oscillation is caused by series and parallel resonance of generalized-impedance of VSCs and grid and thus the small-signal instability can be evaluated by the Nyquist stability criterion. The proposed generalized-impedance also essentially illustrates the stability issues of VSC systems caused by weak grid connection. Digital simulation and RT-LAB simulation of a three-phase grid-connected VSC validates the effectiveness of the proposed GISC.

## II. THE MODEL OF THE GRID-CONNECTED VSC

The grid-connected VSC considered in this paper is shown in Fig. 1. A paralleled LC-type grid impedance is considered, which consists of capacitor $C_f$ and inductance $L_{Line}$. The control system of converter is designed as a PLL based dual-loop vector controller [18, 22-23]. The inner current controller of the VSC has voltage feed-forward (VFF) and decoupling terms. The outer loop can be designed as either output power controller or the dc voltage controller.

This paper focuses on the oscillations introduced by the PLL and the inner/outer control loops. The time delay of the sampling circuits and the PWM modulation are neglected. In addition, based on the time scales of different control loops, the frequency is divided into three parts: medium-low frequency, medium frequency and medium-high frequency, as shown in Table I [24].

To simplify the problem, the resistance of the filter inductance and the transmission lines are neglected, and thus the following assumptions are made in this paper:

1) The power factor of the converter is close to 1 (or -1).
2) The converter connected to the grid is stable. The grid is also stable without the converter. That is, the impedances of the converter and the grid have no unstable poles.

The positions of the phasors and the reference frames when system is suffered from a disturbance are shown in Fig. 2, where the dq frame represents the converter-side rotating frame introduced by the PLL, and the xy frame represents the grid-side synchronous rotating frame [9]. In steady state, the dq frame is synchronized with the xy frame. The angular frequency observed by the PLL is $\omega$. The angle between two frames is $\theta_{pll}$. $\theta_I$ ($\theta_U$) denotes the phase angle of the current (voltage) in the dq frame and $\varphi$ ($\delta$) denotes the phase angel of the current (voltage) in the xy frame. As the medium frequency band can be regarded as the special case of the medium-high frequency band and the medium-low frequency band, we start the analysis firstly from the medium frequency band of the grid-connected converter so as to clearly represent the mathematical deductions and illustrations.

TABLE I. CONVERTER DYNAMICS IN DIFFERENT FREQUENCY BANDS

| Dynamic links | medium-low frequency (few Hertz) | medium frequency (few Hertz ~few tens of Hertz) | medium-high frequency (few tens of Hertz ~few hundreds of Hertz) |
|---|---|---|---|
| reference values for outer loop | × | × | × |
| outer control loop | √ | × | × |
| PLL | √ | √ | √ |
| inner control loop | √ | √ | √ |
| lowpass filter for VFF | × | × | √ |

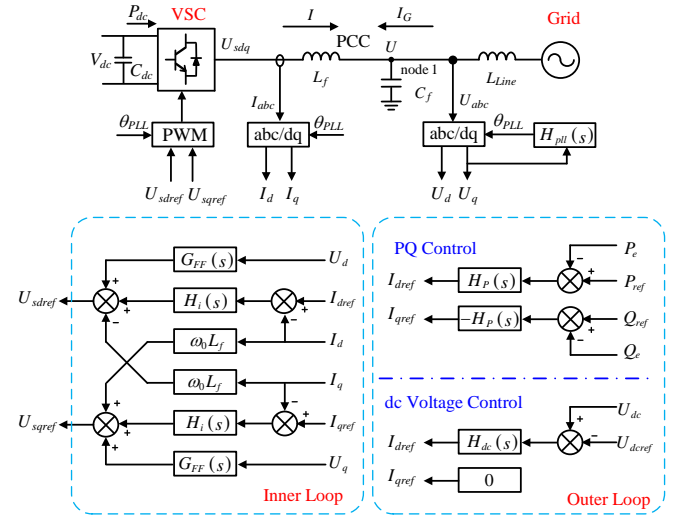

Fig. 1. Block diagram of the grid-connected converter

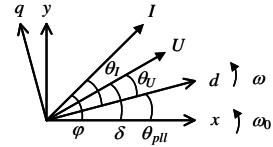

$xy$: grid-side rotating frame (global frame)
$dq$: inverter-side rotating frame (local frame)
$\omega_0$: synchronous speed
$\omega$: speed obtained by PLL

Fig. 2. Reference frames and phasor positions when suffered from a disturbance.

### A. Dynamic Converter Model in Medium-high Frequency Band

The dynamic model of the converter in medium-high and medium frequency bands can be represented by the PLL, current controller, and the filter inductor, as shown in Table I. The dynamic equations of the converter under dq frame can be obtained from Fig.1 and Fig. 2, given as:

$$\begin{cases} \theta_{pll} = H_{pll}(s)U_q \\ \omega = s\theta_{pll} + \omega_0 \end{cases}, \quad (1)$$

$$\begin{cases} U_{sdref} = (I_{dref} - I_d)H_i(s) + G_{FF}(s)U_d - \omega_0 L_f I_q \\ U_{sqref} = (I_{qref} - I_q)H_i(s) + G_{FF}(s)U_q + \omega_0 L_f I_d \end{cases}, \quad (2)$$

$$\begin{cases} U_{sd} = U_d + L_f sI_d - \omega L_f I_q \\ U_{sq} = U_q + L_f sI_q + \omega L_f I_d \end{cases}, \quad (3)$$

where $G_{pll}(s)$ and $G_i(s)$ are the transfer functions of the PLL and the current controller, respectively. $G_{FF}(s)$ is the transfer functions of the lowpass filter used in the voltage feedforward channel. Other symbols are illustrated in Fig. 1 and Fig. 2.

In general, the transfer functions in (1)-(3) are shown as:

$$\begin{cases} G_{FF}(s) = \dfrac{1}{1+sT_{FF}} \\ H_{pll}(s) = \dfrac{1}{s}\left(K_{ppll} + \dfrac{K_{ipll}}{s}\right) \\ H_i(s) = K_{pi} + \dfrac{K_{ii}}{s} \end{cases} \quad (4)$$

Note that the transfer functions may have more complicated expressions in other cases, e.g., when a higher-order filter is used. However, as the following mathematical derivation is carried out based on the commonly used symbols (i.e. $G_{FF}(s)$, $H_{pll}(s)$, and $H_i(s)$), this analysis method is still applicable on other complex cases.

To obtain the dynamic model for small-signal analysis of the converter, the following steps are conducted based on the dynamic equations (1)-(3) : 1) The small-signal model in the dq frame is obtained by linearizing (1)-(3) and solving the linearized equations; 2) The model in the dq frame is rewritten in polar form; 3) Such model is transformed to the xy frame using the angle relationships as shown in Fig. 2. The details are given in Appendix A. Then the small-signal model of the converter in the xy frame can be written as:

$$\begin{bmatrix} \Delta I \\ I\Delta\varphi \end{bmatrix} = \begin{bmatrix} Y_{g1}(s) & Y_{g2}(s) \\ Y_{g3}(s) & Y_{g4}(s) \end{bmatrix} \begin{bmatrix} \Delta U \\ U\Delta\delta \end{bmatrix}, \quad (5)$$

where

$$\begin{cases} Y_{g1}(s) = \dfrac{1-G_{FF}(s)}{sL_f + H_i(s)}\cos\theta_I \\ Y_{g2} = \dfrac{1-G_{FF}(s)}{sL_f + H_i(s)}\dfrac{1}{1+H_{pll}(s)U_0}\sin\theta_I \\ Y_{g3} = -\dfrac{1-G_{FF}(s)}{sL_f + H_i(s)}\sin\theta_I \\ Y_{g4} = \dfrac{(1-G_{FF}(s))\cos\theta_I - H_i(s)H_{pll}(s)I_0}{(sL_f + H_i(s))(1+H_{pll}(s)U_0)} \end{cases}.$$

The subscript "0" denotes steady-state values.

Since assumption 1 yields $\cos\theta_I \approx 1$ and $\sin\theta_I \approx 0$, the value of $Y_{g2}$ and $Y_{g3}$ are far less than $Y_{g1}$ and approximately equal to 0. Therefore, (5) can be approximated to:

$$\begin{bmatrix} \Delta I \\ I\Delta\varphi \end{bmatrix} = \begin{bmatrix} Y_{g1} & 0 \\ 0 & Y_{g4} \end{bmatrix} \begin{bmatrix} \Delta U \\ U\Delta\delta \end{bmatrix}. \quad (6)$$

Especially, as $G_{FF}(s) \approx 1$ is satisfied in the medium frequency band, (5) can be simplified as:

$$\begin{bmatrix} \Delta I \\ I\Delta\varphi \end{bmatrix} = \begin{bmatrix} 0 & 0 \\ 0 & Y_{g4} \end{bmatrix} \begin{bmatrix} \Delta U \\ U\Delta\delta \end{bmatrix}, \quad (7)$$

where:

$$Y_{g4} = \dfrac{H_i(s)}{sL_f + H_i(s)}\dfrac{H_{pll}(s)I_0}{1+H_{pll}(s)U_0}.$$

It can be seen from (5) that $|H_{pll}(s)|$ gets smaller with the increase of frequency, which makes $Y_{g1}$ and $Y_{g4}$ tend to be equal. Especially, when the frequency is high enough, $H_{pll}(s) \approx 0$ and $Y_{g1} \approx Y_{g4}$ are satisfied.

*B. Dynamic Converter Model in Medium-low Frequency Band*

When the oscillation frequency is relatively low, the system stability is also affected by dynamics of the outer control loop. Therefore, the small-signal model of the converter should also take into consideration the outer control loop. Two typical controllers, i.e., the PQ controller and the dc voltage controller, are considered in the following analysis, as shown in Fig.1.

1) PQ control

The equations of the power controller and the power measurement should be additionally added to the dynamic equation (1)-(3), expressed as:

$$\begin{cases} I_{dref} = (P_{ref} - P_e)H_P(s) \\ I_{qref} = -(Q_{ref} - Q_e)H_P(s) \end{cases}, \quad (8)$$

$$\begin{cases} P_e = G_P(s)(U_d I_d + U_q I_q) \\ Q_e = G_P(s)(-U_d I_q + U_q I_d) \end{cases}, \quad (9)$$

where $H_P(s)$ is the transfer function of the power control loop, $P_{ref}$ and $Q_{ref}$ are the reference values of the active power and the reactive power, $G_P(s)$ is the equivalent transfer function of the power measurement link. In general, the transfer functions in (8)-(9) are shown as:

$$\begin{cases} G_P(s) = \dfrac{1}{1+sT_P} \\ H_P(s) = K_{pp} + \dfrac{K_{ip}}{s} \end{cases}.$$

Linearizing (1)-(3), (8)-(9) and transforming them to xy frame, the model of the converter can be derived as (10). The detailed derivation process is shown in Appendix A.

$$\begin{bmatrix} \Delta I \\ I\Delta\varphi \end{bmatrix} = \begin{bmatrix} Y_{g1} & 0 \\ 0 & Y_{g4} \end{bmatrix} \begin{bmatrix} \Delta U \\ U\Delta\delta \end{bmatrix}, \quad (10)$$

where

$$\begin{cases} Y_{g1} = \dfrac{H_P(s)H_i(s)G_P(s)I_0}{H_P(s)H_i(s)G_P(s)U_0 + H_i(s) + L_f s} \\ Y_{g4} = \dfrac{sL_f H_{pll}I_0 - H_P H_i G_P I_0}{(H_P H_i G_P U_0 + H_i + L_f s)(1+U_0 H_{pll})} - \dfrac{I_0 H_{pll}}{1+U_0 H_{pll}} \end{cases}.$$

2) dc voltage control

When the dc voltage control is adopted for outer loop, the equation of the dc voltage control loop (11) and the dynamics of the dc capacitor (12) should be added to the dynamic equation (1)-(3) to describe converter dynamic characteristics in this case,

$$I_{dref} = (U_{dc} - U_{dcref})H_{dc}(s), \quad (11)$$

$$U_{dc}C_{dc}\dfrac{dU_{dc}}{dt} = -P_e + P_m, \quad (12)$$

where $U_{dcref}$ ($U_{dc}$) is the reference (measured) value of the dc voltage. $P_m$ ($P_e$) is the input (output) power of the dc bus. $H_{dc}(s)$ is the transfer function of the dc voltage controller, which is further defined as:

$$H_{dc}(s) = K_{pdc} + \dfrac{K_{idc}}{s}.$$

Similarly, linearizing (1)-(3) and (11)-(12) and transforming them to xy frame, the model of the converter can be obtained as (13), with the formula derivation process given in Appendix A for details.

$$\begin{bmatrix} \Delta I \\ I\Delta\varphi \end{bmatrix} = \begin{bmatrix} Y_{g1} & 0 \\ 0 & Y_{g4} \end{bmatrix} \begin{bmatrix} \Delta U \\ U\Delta\delta \end{bmatrix}, \quad (13)$$

where

$$\begin{cases} Y_{g1} = -\dfrac{H_{dc}(s)H_i(s)I_0}{sU_{dc0}C_{dc}(H_i(s)+sL_f)+H_{dc}(s)H_i(s)U_0} \\ Y_{g4} = \dfrac{H_i(s)H_{pll}(s)I_0}{(H_i(s)+sL_f)(1+H_{pll}(s)U_0)} \end{cases}.$$

Note that this method can be also applied to describe a direct driven wind turbine generator or a photovoltaic generator, where $P_m$ is regarded as the input power. To analyze the impact of the dynamics of the input power $P_m$ on system stability, $P_m$ can be modeled as the dynamics of the dc bus, in which the equation (13) is still adaptable.

### C. Dynamic Model of AC Grid

According to the topology in Fig.1, the grid side circuit equation can be modeled using the linearized state equation of the inductor and the capacitor. The linearized state equation of the transmission line inductor in xy frame is given as:

$$\begin{bmatrix} \Delta I_{kx} \\ \Delta I_{ky} \end{bmatrix} = \dfrac{1}{L_k s^2 + L_k \omega^2}\begin{bmatrix} s & \omega \\ -\omega & s \end{bmatrix}\left\{\begin{bmatrix} \Delta U_{ix} \\ \Delta U_{iy} \end{bmatrix} - \begin{bmatrix} \Delta U_{jx} \\ \Delta U_{jy} \end{bmatrix}\right\}, \quad (14)$$

where $i$ and $j$ denote the node next to the inductor.

Since the voltage of the ideal grid is considered to be constant, the linearized state equation of $L_{Line}$ can be expressed as:

$$\begin{bmatrix} \Delta I_{Linex} \\ \Delta I_{Liney} \end{bmatrix} = \dfrac{1}{L_{Line}s^2 + L_{Line}\omega^2}\begin{bmatrix} s & \omega \\ -\omega & s \end{bmatrix}\begin{bmatrix} \Delta U_x \\ \Delta U_y \end{bmatrix}. \quad (15)$$

Denote $\phi_{Line}$ to be the power factor angle at the node 1 of $L_{Line}$. Then (15) can be converted to the polar coordinates as follows:

$$\begin{bmatrix} \Delta I_{Line} \\ I_{Line}\Delta\varphi_{Line} \end{bmatrix} = \mathbf{Y}_{\mathbf{Line}}\begin{bmatrix} \Delta U \\ U\Delta\delta \end{bmatrix}, \quad (16)$$

where $\mathbf{Y}_{\mathbf{Line}} = \dfrac{1}{L_{Line}s^2 + L_{Line}\omega^2}\begin{bmatrix} s & \omega \\ -\omega & s \end{bmatrix}\begin{bmatrix} \cos\phi_{Line} & -\sin\phi_{Line} \\ \sin\phi_{Line} & \cos\phi_{Line} \end{bmatrix}$.

Similarly, the linearized state equation of a capacitor in xy frame can be expressed as:

$$\begin{bmatrix} \Delta I_{kx} \\ \Delta I_{ky} \end{bmatrix} = C_k\begin{bmatrix} s & -\omega \\ \omega & s \end{bmatrix}\left\{\begin{bmatrix} \Delta U_{ix} \\ \Delta U_{iy} \end{bmatrix} - \begin{bmatrix} \Delta U_{jx} \\ \Delta U_{jy} \end{bmatrix}\right\}, \quad (17)$$

where $i$ and $j$ denote the node next to the capacitor.

Denote $\phi_C$ to be the power factor angle of $C_f$ at the node 1. Since the voltage of the ground is also constant, the linearized state equation of $C_f$ can be expressed as:

$$\begin{bmatrix} \Delta I_C \\ I_C\Delta\varphi_C \end{bmatrix} = \mathbf{Y}_{\mathbf{C}}\begin{bmatrix} \Delta U \\ U\Delta\delta \end{bmatrix}, \quad (18)$$

where $\mathbf{Y}_{\mathbf{C}} = C_C\begin{bmatrix} s & -\omega \\ \omega & s \end{bmatrix}\begin{bmatrix} \cos\phi_C & -\sin\phi_C \\ \sin\phi_C & \cos\phi_C \end{bmatrix}$.

According to the circuit topology shown in Fig. 1, the model of the network can be formulated as follows:

$$\Delta \mathbf{I_G} = \begin{bmatrix} \Delta I_G \\ I_G\Delta\varphi_G \end{bmatrix} = -\Delta \mathbf{I} = -\mathbf{Y}\begin{bmatrix} \Delta U \\ U\Delta\delta \end{bmatrix}, \quad (19)$$

where $\mathbf{Y} = \mathbf{Y}_{\mathbf{Line}} + \mathbf{Y}_{\mathbf{C}}$. In particular, $\mathbf{Y}_{\mathbf{C}}$ is 0 if the output filter only has an inductor.

## III. Generalized-impedances Based Stability Criterion

### A. The Generalized-impedance of the Grid-connected Converter System

It can be seen from (19) that the admittance matrix of the ac grid can be written as the form $\begin{bmatrix} A(s) & -C(s) \\ C(s) & A(s) \end{bmatrix}$.

It can be observed from (6), (7), (10), and (13) that the admittance matrix of the converter can be written as the form $\begin{bmatrix} A(s) & 0 \\ 0 & D(s) \end{bmatrix}$. Consequently, the dynamic model of the converter and the grid can be written as the common expression as follows:

$$\begin{bmatrix} \Delta I \\ I\Delta\varphi \end{bmatrix} = \begin{bmatrix} A(s) & -C(s) \\ C(s) & D(s) \end{bmatrix}\begin{bmatrix} \Delta U \\ U\Delta\delta \end{bmatrix}. \quad (20)$$

Using (19) and (20), the characteristic equation of the whole system can be written as follows:

$$\det\left(\begin{bmatrix} Y_{g1}(s) & \\ & Y_{g4}(s) \end{bmatrix} + \begin{bmatrix} A(s) & -C(s) \\ C(s) & A(s) \end{bmatrix}\right) = 0, \quad (21)$$

where det( ) is the determinant of the matrix.

**Definition 1** (Generalized-admittance): For a dynamic model shown as (20), its generalized-admittances $Y_{ei}(s)\,(i=1,2,3)$ are defined as:

$$\begin{cases} Y_{e1}(s) := D(s) - A(s) \\ Y_{e2}(s) := A(s) + jC(s) \\ Y_{e3}(s) := A(s) - jC(s) \end{cases}. \quad (22)$$

The generalized-impedance is defined as the inverse of the generalized-admittance, i.e., $Z_{ei}(s) = Y_{ei}^{-1}(s)$.

According to the definition and the model of the converter (5) and (10), the generalized-admittance of the converter can be written as:

$$\begin{cases} Y_{e1\_VSC}(s) = Y_{g4}(s) - Y_{g1}(s) \\ Y_{e2\_VSC}(s) = Y_{g1}(s) \\ Y_{e3\_VSC}(s) = Y_{g1}(s) \end{cases}. \quad (23)$$

To obtain the generalized-impedances of the grid and to analyze the characteristic of the admittance matrix of the grid, the following transformation matrices are defined:

$$\begin{cases} \mathbf{T} = \dfrac{1}{\sqrt{2}}\begin{bmatrix} 1 & j \\ 1 & -j \end{bmatrix} \\ \mathbf{T}^{-1} = \dfrac{1}{\sqrt{2}}\begin{bmatrix} 1 & 1 \\ -j & j \end{bmatrix} \end{cases}, \quad (24)$$

where $j = \sqrt{-1}$.

Since matrix $\mathbf{Y}$ has the same structure as $\begin{bmatrix} a & -b \\ b & a \end{bmatrix}$, and $\mathbf{T}\begin{bmatrix} a & -b \\ b & a \end{bmatrix}\mathbf{T}^{-1} = \begin{bmatrix} a+jb & 0 \\ 0 & a-jb \end{bmatrix}$, $\mathbf{TYT^{-1}}$ is also a diagonal matrix, which can be written as

$$\mathbf{TYT^{-1}} = \begin{bmatrix} Y_+ & \\ & Y_- \end{bmatrix} = \begin{bmatrix} Z_+^{-1} & \\ & Z_-^{-1} \end{bmatrix}. \quad (25)$$

According to the definition, the generalized-admittance of the grid can be written as:

$$\begin{cases} Y_{e1\_grid}(s) = 0 \\ Y_{e2\_grid}(s) = Y_+ \\ Y_{e3\_grid}(s) = Y_- \end{cases}. \quad (26)$$

### B. Characteristic Equation and the Generalized-impedance Based Stability Criterion in Medium Frequency Band

In medium frequency band, considering the converter side output equation (7) and grid side output equation (19), the following equation is established

$$0 = (\begin{bmatrix} 0 & 0 \\ 0 & Y_{g4} \end{bmatrix} + \mathbf{Y})\begin{bmatrix} \Delta U \\ U\Delta\delta \end{bmatrix}. \quad (27)$$

Therefore, the characteristic equation (21) is simplified as,

$$\det(\begin{bmatrix} 0 & 0 \\ 0 & Y_{g4} \end{bmatrix} + \mathbf{Y}) = 0. \quad (28)$$

Since matrix $\mathbf{T}$ is invertible, the following equivalence holds:

$$\det[\mathbf{T}(\begin{bmatrix} 0 & 0 \\ 0 & Y_{g4} \end{bmatrix} + \mathbf{Y})\mathbf{T}^{-1}] = 0 \Leftrightarrow$$
$$\det(\frac{Y_{g4}}{2}\begin{bmatrix} 1 & -1 \\ -1 & 1 \end{bmatrix} + \begin{bmatrix} Y_+ & 0 \\ 0 & Y_- \end{bmatrix}) = 0. \quad (29)$$

Then the equation (29) is discussed from the following two aspects:

1) It can be found from (29) that $\det(Y_+(s)Y_-(s)) = 0$ if $Y_{g4}(s) = 0$, and vice versa, i.e., $Y_{g4}(s) = 0$, if $\det(Y_+(s)Y_-(s)) = 0$. In this special case that $Y_{g4}(s) = 0$ and $\det(Y_+(s)Y_-(s)) = 0$ are satisfied simultaneously, the grid and the converter are both unstable and have the same resonance point. Due to the rarity of this special case, it is not the focus of this paper.

2) Except for the above case, $\det(Y_+(s)Y_-(s)) \neq 0$ and $Y_{g4}(s) \neq 0$ hold simultaneously. Given that, (29) is equivalent to

$$\det(\begin{bmatrix} Z_+ & -Z_- \\ -Z_+ & Z_- \end{bmatrix} + 2Z_{g4}\mathbf{I})\det(\frac{Y_{g4}\mathbf{I}}{2}\begin{bmatrix} Y_+ & \\ & Y_- \end{bmatrix}) = 0$$
$$\Leftrightarrow \det[\begin{bmatrix} 1 & 0 \\ 1 & 1 \end{bmatrix}(\begin{bmatrix} Z_+ & -Z_- \\ -Z_+ & Z_- \end{bmatrix} + 2Z_{g4}\mathbf{I})\begin{bmatrix} 1 & 0 \\ -1 & 1 \end{bmatrix}\frac{Y_{g4}\mathbf{I}}{2}] = 0$$
$$\Leftrightarrow \det[\begin{bmatrix} 2Z_{g4}+(Z_++Z_-) & -Z_- \\ 0 & 2Z_{g4} \end{bmatrix}\frac{Y_{g4}\mathbf{I}}{2}] = 0$$
$$\Leftrightarrow \det[\begin{bmatrix} 1+\frac{Y_{g4}(Z_++Z_-)}{2} & -\frac{Y_{g4}Z_-}{2} \\ 0 & 1 \end{bmatrix}] = 0 \quad . \quad (30)$$

It follows from (30) that the characteristic equation of the closed-loop system can be simplified as

$$1 + \frac{Y_{g4}(Z_++Z_-)}{2} = 0. \quad (31)$$

Equation (31) describes the stability with the interaction of the grid and the converter. The characteristic roots of (31) are the closed-loop poles of the system. According to the relationship between the positions of the poles and the system stability, the necessary condition of a stable system is that all the characteristic roots of (31) should lie on the left half of the complex plane. Since (31) can be also regarded as the characteristic equation of a SISO system, the Nyquist criterion can be used to detect the unstable poles of the system. $Y_{g4}(s)(Z_+(s)+Z_-(s))/2$ can be regarded as the open-loop transfer function of the SISO system. Further, assumption 2 implies that the converter is stable when connected to an ideal grid ($Y_{g4}(s)$ has no unstable poles). It also implies that $(Z_+(s)+Z_-(s))$ has no unstable poles. Consequently, it obeys the Nyquist criterion that

**GISC 1**: If the Nyquist curve of $Z_{G\_grid}(s)/Z_{G\_VSC}(s)$ (or $Y_{g4}(s)Z_{G\_grid}(s)$) does not encircle the point $(-1, j0)$, the system is stable, where $Z_{G\_grid}(s) = (Z_+(s)+Z_-(s))/2$, $Z_{G\_VSC}(s) = Z_{g4}(s) = 1/Y_{g4}(s)$.

Since the distance from the point $(-1, j0)$ to the Nyquist curve can reflect the stability and stability margin of the system, (31) can be also used to guide oscillation suppression. Moreover, as the system represented by (31) is a SISO system rather than the original two-input-two-output system, the system stability analysis can be simplified. It is the major difference compared with the traditional impedance base stability analysis for a three-phase system.

### C. Physical Interpretation for the Generalized-Impedance Based Stability Criterion in Medium Frequency Band

The stability of the system can be analyzed through the characteristic equation (31), which can be rewritten as

$$Z_{g4}(s) + Z_+(s) + Z_{g4}(s) + Z_-(s) = 0, \quad (32)$$

or written in terms of the generalized-impedances due to $Y_{g1} = 0$, as follows

$$Z_{e1\_VSC} + Z_{e2\_grid} + Z_{e1\_VSC} + Z_{e3\_grid} = 0. \quad (33)$$

It can be observed that the characteristic equation describing the converter system can be interpreted as an equivalent circuit composed of the generalized-impedances of the converter and the grid. The equivalent circuit is shown in Fig. 3 (a), where the dashed line denotes an infinite impedance.

In terms of the physical meaning, the oscillation in a three-phase system is equivalent to the resonance of the equivalent circuit. If the oscillation frequency of the equivalent circuit is $f_R$, the oscillation frequencies in the original system are $f_0 \pm f_R$. The physical system implies that the system is unstable when the total resistance of the generalized-impedances is negative.

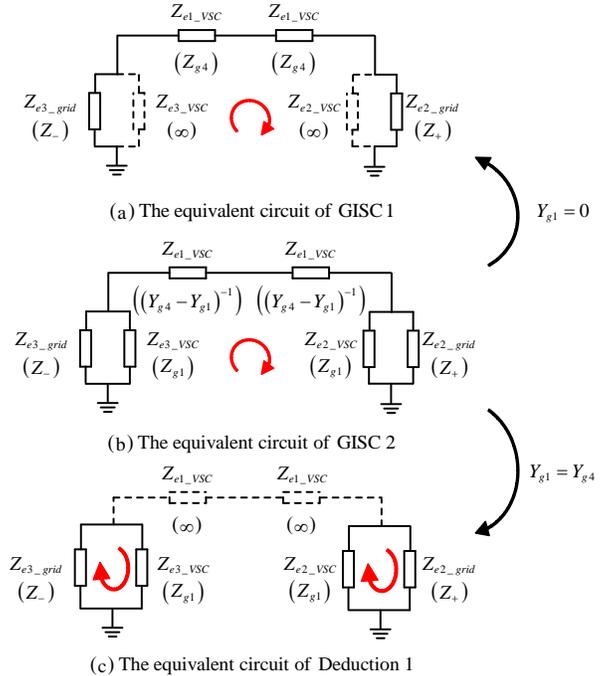

Fig. 3 Equivalent circuit of the system

The presented method has potential to be applied for analysis of complex system with multi-converters, e.g. a system with N converters as shown in Fig.4. To describe a multi-converter system, the characteristic equation (30) will be slightly adjusted, where the generalized-admittance of the converter becomes a diagonal matrix and the generalized-impedance of the grid is written as $Z_+ + Z_-$.

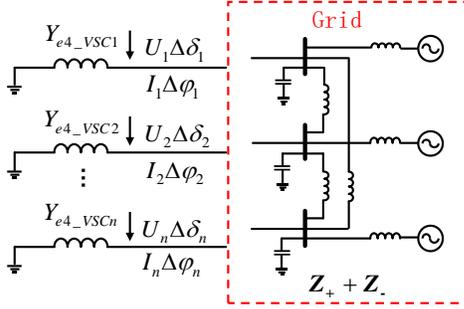

Fig. 4 Schematic diagram of multiple-VSC system

### D. The Generalized-Impedance Based Stability Criterion in Full Frequency Band

It can be observed from (6), (7), (10), and (13) that the common form can be formulated to describe the dynamic model of the converter in all frequency bands considered in this paper, given as follows:

$$\begin{bmatrix} \Delta I \\ I\Delta\varphi \end{bmatrix} = \begin{bmatrix} Y_{g1} & 0 \\ 0 & Y_{g4} \end{bmatrix} \begin{bmatrix} \Delta U \\ U\Delta\delta \end{bmatrix}. \quad (34)$$

The characteristic equation of the closed-loop system can be derived from equation (34) and the dynamic equation of the grid (19), given as follows:

$$\det(\begin{bmatrix} Y_{g1} & \\ & Y_{g4} \end{bmatrix} + \mathbf{Y}) = 0. \quad (35)$$

Equation (35) can be transformed to

$$\det(\begin{bmatrix} 0 & 0 \\ 0 & Y_{g4}-Y_{g1} \end{bmatrix} + \begin{bmatrix} Y_{g1} & 0 \\ 0 & Y_{g1} \end{bmatrix} + \mathbf{Y}) = 0 \Leftrightarrow$$

$$\det[\mathbf{T}(\begin{bmatrix} 0 & 0 \\ 0 & Y_{g4}-Y_{g1} \end{bmatrix} + \begin{bmatrix} Y_{g1} & 0 \\ 0 & Y_{g1} \end{bmatrix} + \mathbf{Y})\mathbf{T}^{-1}] = 0. \quad (36)$$

Taking $Y_{g1}$ as a part of the admittance matrix of the grid $\mathbf{Y}$ and using the same procedure as (30), we can transform the characteristic equation of the system to the following equation,

$$1 + (Y_{g4} - Y_{g1})\frac{(Y_+ + Y_{g1})^{-1} + (Y_- + Y_{g1})^{-1}}{2} = 0, \quad (37)$$

which can be also written as the following form

$$2(Y_{g4} - Y_{g1})^{-1} + (Y_+ + Y_{g1})^{-1} + (Y_- + Y_{g1})^{-1} = 0. \quad (38)$$

According to the definition of the generalized-impedance, the characteristic equation (37) or (38) can be also interpreted as an equivalent circuit composed of the generalized-impedances of the grid and the converter, as shown in Fig. 3 (b). The oscillation of the system is equivalent to the resonance of the equivalent circuit.

Denote $(Y_+ + Y_{g1})^{-1}$ as $Z'_+$, $(Y_- + Y_{g1})^{-1}$ as $Z'_-$, $(Z'_+ + Z'_-)/2$ as $Z'_{G\_grid}$, and $(Y_{g4} - Y_{g1})^{-1}$ as $Z'_{G\_VSC}$, the stability criterion for the whole frequency band can be obtained.

**GISC 2**: Given that the Nyquist curve of $Z'_{G\_grid}(s)/Z'_{G\_VSC}(s)$ (or $Y'_{G\_VSC}(s)Z'_{G\_grid}(s)$) anti-clockwise encircles the point $(-1, j0)$ for M times, if M equals the number of the unstable poles of $Z'_{G\_grid}(s)$, the system is stable.

Note that (37) can be rewritten as

$$(Z'_{G\_grid}(s) + Z'_{G\_VSC}(s))Z_{g1} + \Delta(s) = 0, \quad (39)$$

where $\Delta(s) = Z_+Z_- + Z_{g4}(Z_+ + Z_-)/2$.

Especially, as $Y_{g1} \approx 0$ ($Z_{g1}$ tends to infinity) holds in the medium frequency band, $\Delta(s)$ will be much less than $(Z'_{G\_grid}(s) + Z'_{G\_VSC}(s))Z_{g1}$. Therefore, the solution of (39) can be approximately taken as the solution of $Z'_{G\_grid}(s) + Z'_{G\_VSC}(s) = 0$. In this scenario, the stability can be analyzed using GISC 1. Consequently, GISC 1 can be regarded as a special case of GISC 2 when $Y_{g1} = 0$.

When $Y_{g4}$ and $Y_{g1}$ tend to be equal, (37) implies that $(Y_+ + Y_{g1})^{-1} + (Y_- + Y_{g1})^{-1}$ tends to infinity and thus $Y_+ + Y_{g1}$ and $Y_- + Y_{g1}$ tend to be 0. Therefore, (37) is equivalent to:

$$(Y_+ + Y_{g1})(Y_- + Y_{g1}) = 0. \quad (40)$$

Similar to the aforementioned analysis, the stability of the system can be analyzed through (40) using Nyquist criterion. In other words, when the difference between the value of $Y_{g4}$ and $Y_{g1}$ is small enough, GISC 2 is equivalent to the following conclusion:

**Deduction 1**: If the Nyquist curve of $Z_{g1}(s)/Z_+(s)$ and $Z_{g1}(s)/Z_-(s)$ do not encircle the point $(-1, j0)$, the system is stable.

Note that the stability criterion concluded from the deduction 1 can be also obtained by the analysis in sequence domain. Given that $Y_{g4} \approx Y_{g1}$, the GISC is equivalent to the traditional impedance based stability criterion. The equivalent circuits of the GISC 1, GISC 2, and the deduction 1 are shown in Fig. 3.

Note that the previous analysis is based on a general characteristic equation expressed by (21). In essence, the analysis transforms a MIMO system with a certain special structure to an equivalent SISO system. Such analysis method is neither limited by the reference frame used to model the system nor the control strategy of the converter. Given that the impedance matrix of the converter is diagonal and the impedance matrix of the grid has the same structure as $\begin{bmatrix} a & -b \\ b & a \end{bmatrix}$, the presented analysis method and the GISCs can be used in more scenarios and have huge potential for practical applications.

## IV. CASE STUDY AND EXPERIMENTAL VERIFICATION

In this section, case studies and experiments are carried out to verify the effectiveness of the proposed analysis method. The grid-connected converter and its control strategy are shown in Fig.1. The voltage of the dc bus is supposed to be constant. The reference values for the current controller are $I_{dref} = 1.0pu$ and $I_{qref} = 0pu$. An inductor at converter side is used for harmonic damping. Other parameters are listed in Table II.

### A. The Mechanism of the Weak-Grid-Caused Instability

Engineering experience shows that the oscillation problem gets serious when the grid turns weak. In this subsection, we use the proposed GISC and the corresponding analysis method to reveal the mechanism of this phenomenon.

According to (16), the impedances of the grid on the condition that $\cos\phi = 1$ can be shown as:

$$\begin{cases} Z_+ = (L_{\text{line}}s + jL_{\text{line}}\omega)(\cos\phi - j\sin\phi) = L_{\text{line}}s + jL_{\text{line}}\omega \\ Z_- = (L_{\text{line}}s - jL_{\text{line}}\omega)(\cos\phi + j\sin\phi) = L_{\text{line}}s - jL_{\text{line}}\omega \end{cases}.$$

To better uncover the cause of instability, the filter in the voltage feedforward is neglected. The characteristic equation of the system can be written as $sL_{\text{line}} + Z_{g4} = 0$, which is equivalent to:

$$\frac{1}{L_{\text{line}}} + sY_{g4} = 0. \tag{41}$$

When s increases from $-j\infty$ to $j\infty$ along the $j\omega$ axis, the Nyquist curve of $-sY_{g4}$ encircling $(1/L_{\text{line}}, j0)$ is equivalent to the Nyquist curve of $Z_{G\_line}(s)/Z_{G\_VSC}(s)$ encircling $(-1, j0)$. Denote $-sY_{g4}$ as the open-loop transfer function $G_{OL}(s)$, its Nyquist curve is shown in Fig.6.

It can be seen from Fig.6 that the Nyquist curve of $G_{OL}(s)$ goes clockwise for 2 laps. According to GISC 1, if the curve encircles $(1/L_{\text{line}}, j0)$, the system has two poles on the right half plane, making the system unstable. When the length of the transmission line is small and $1/L_{\text{line}}$ is relatively big, the point $(1/L_{\text{line}}, j0)$ locates outside the unstable area, e.g. point A in Fig.6, leading to a stable system. With the increase of the length of transmission line, the point $(1/L_{\text{line}}, j0)$ moves to the left and the point $(1/L_{\text{line}}, j0)$ will fall into the unstable area, e.g., point B in Fig.6, leading to an unstable system. It can be concluded that the proposed GISC has good performance to illustrate the relationship between the oscillation and the weakness of the connected grid.

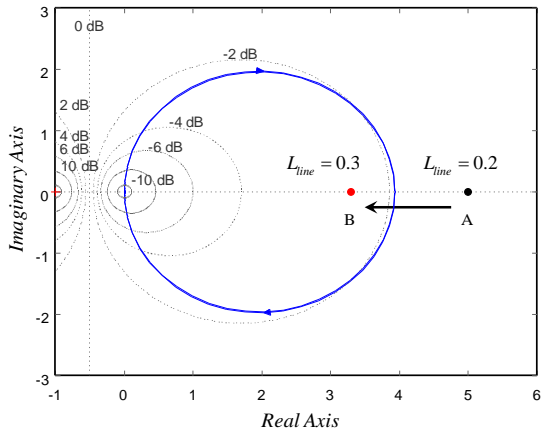

Fig. 6 Instability caused by weak grid

### B. GISC Based Stability Analysis of the System in Critical Scenarios

The stability analysis of the system in two critical scenarios when the line inductance equals 0.20pu and 0.26pu are carried out using GISC based analysis method, with the corresponding Nyquist curve of $Z'_{G\_grid}(s)/Z'_{G\_VSC}(s)$ in Fig. 7. It can be seen that the Nyquist curve are two clockwise circles and intersect the real axis with two points. When the inductance is 0.2pu, the left point is on the right side of $(-1, j0)$. Thus, the Nyquist curve doesn't encircle the point $(-1, j0)$ and the system is stable. When the inductance is 0.26pu, the left point is on the left side of $(-1, j0)$. Therefore, the Nyquist curve clockwise encircles the point $(-1, j0)$ for two times and the system has two poles on the right half plane, leading to an unstable system. The root locus depicted in Fig.8 also draws the same conclusion.

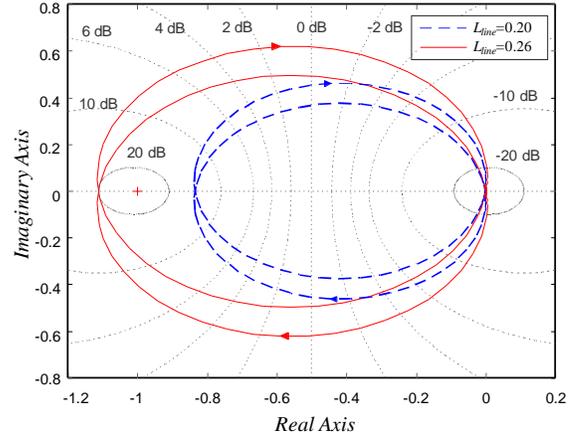

Fig. 7 Nyquist plot with different line inductance

The root locus in Fig. 8 also shows that the inductance' threshold value to system instability is around 0.23pu, where two poles on the imaginary axis reflect the undamped oscillation in the system. When the inductance is 0.2pu and 0.26pu, the predominant poles locate at the left side and right side of the imaginary axis respectively, but both very close to the imaginary axis. Hence, the accuracy of GISC method in distinguishing the two contiguous cases can be verified.

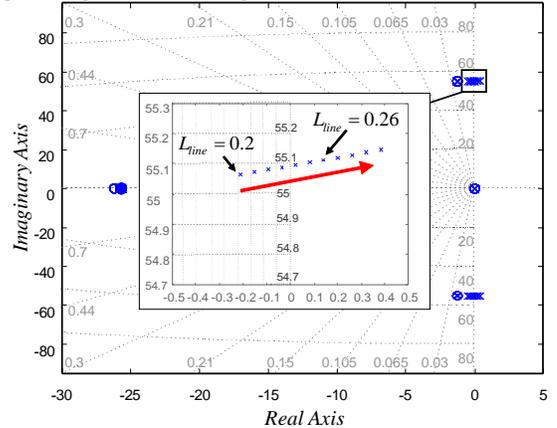

Fig. 8 Root locus when line inductance increases

### C. Simulation Results

To further verify the proposed criterion, the Matlab/SIMULINK based digital simulation and the RTLAB based hardware-in-the-loop (HIL) simulation are carried out in this subsection. For Matlab/SIMULINK simulation, the control frequency is set as 4kHz, the SVPWM frequency is set as 4kHz, the step interval of the simulation is set as $5\mu s$.

To study the effect of the line inductance on system stability, the line inductance is as set to step from 0.20pu to 0.26pu (or the critical value 0.234pu) at $t = 5s$. The waveforms of the voltage and the current given in dq frame before and after the change of the line inductance are shown in Fig. 9 and Fig. 10 respectively. The spectrum of the oscillating voltage is shown in Fig. 11.

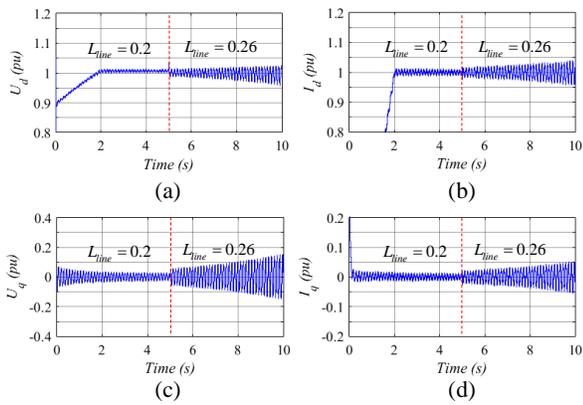

Fig. 9 Voltage and current waveforms after the inductance is changed to 0.26pu

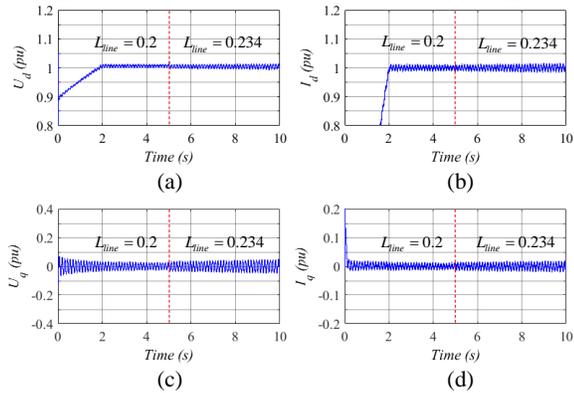

Fig. 10 Voltage and current waveforms after the inductance is changed to the critical value

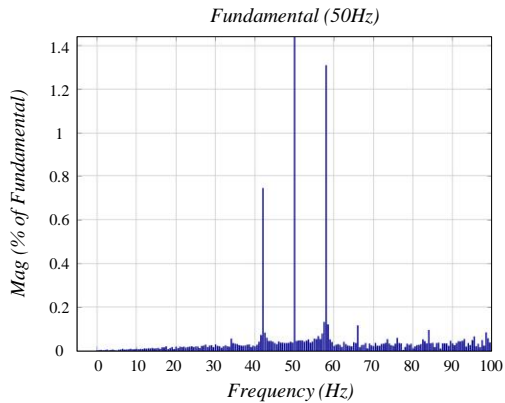

Fig. 11 Frequency spectrum of converter current

It can be found from Fig. 9 and Fig. 10 that the oscillation is damped when the line inductance is 0.20pu. When the line inductance is larger than the critical value, e.g., 0.26pu, the oscillation is divergent. Fig. 11 shows that the oscillation frequencies are 42Hz and 58Hz respectively, when the system is marginally stable. Hence, the aforementioned analysis can be verified by the simulation results.

The hardware-in-the-loop (HIL) simulation based on RTLAB also verifies the analysis results. The converter and the grid are simulated by OP5600 simulator. The controller of the converter is based on DSP TMS320F28335. The control frequency is set as 4kHz. The frequency of the SVPWM is 4kHz. The HIL test platform is shown in Fig. 12.

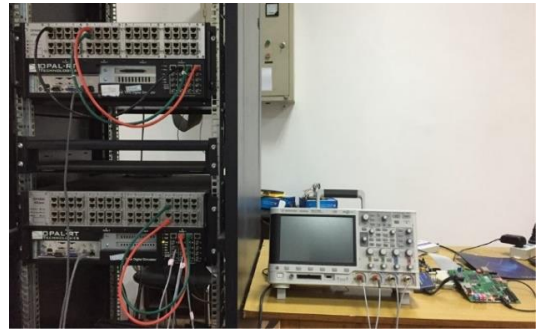

Fig. 12 HIL test platform based on RTLAB

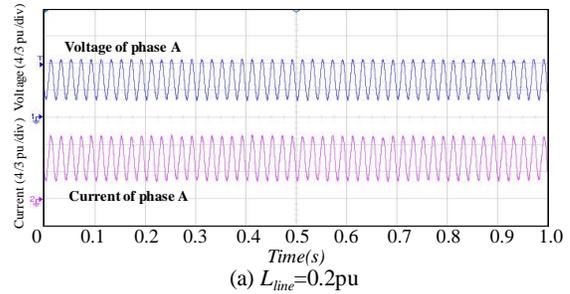

(a) $L_{line}$=0.2pu

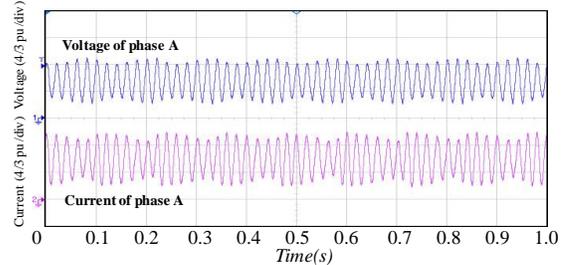

(b) $L_{line}$=0.26pu

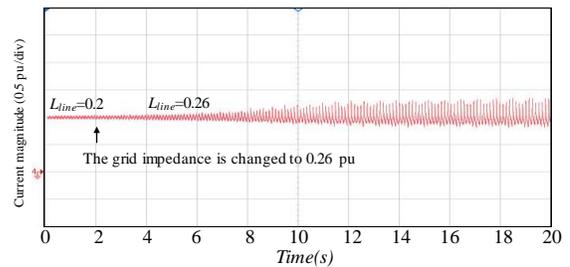

(c) Magnitude of the current

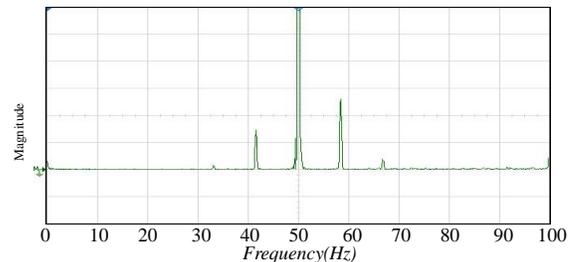

(d) Frequency spectrum of the voltage

Fig. 13 Voltage and current responses when $L_{line}$ steps from 0.20pu to 0.26pu

Similarly, the inductance is assumed to step from 0.2pu to 0.26pu at $t = 2s$. The voltage and the current responses in this condition are shown in Fig. 13. The voltage and the current waveforms of phase A with different line inductances are shown in Fig. 13(a) and (b). The magnitude of the output current is shown in Fig. 13(c). The frequency spectrum of the voltage of phase A is shown in Fig. 13(d). The currents and the voltages are all per-unit values based on the rated power and voltage of the system. Fig. 13 shows that the oscillation phenomenon and the oscillation frequency (42Hz and 58Hz), can be observed, which proof that both digital and HIL simulation results are consistent with each other.

TABLE II. PARAMETERS OF THE VSC

| Symbol | Description | Value |
| --- | --- | --- |
| $S_b$ | Base value of power | 500kVA |
| $U_b$ | Base value of voltage | 690V |
| $L_f$ | Inductance of the inverter side filter | 0.2pu |
| $G_i(s)$ | Transfer function of the current controller | $0.6+15/s$ |
| $G_{pll}(s)$ | Transfer function of the PLL | $2.5/s+3020/s^2$ |

## V. CONCLUSION

The generalized-impedance based stability criterion is innovatively proposed for the stability analysis of a three-phase grid-connected VSC system. Through strict mathematical derivation, a MIMO system composed of the VSC and the grid is transformed to an equivalent SISO system composed of the generalized-impedances of the grid and the VSC. Consequently, the oscillation can be regarded as the series resonance of the equivalent circuit. Then the GISC is proposed for the stability analysis of a three-phase system. GISC based stability criterion well take into consideration the significant effects of the PLL dynamics on system stability. Digital simulation and HIL simulation validate the effectiveness of the proposed criterion. For future work, this criterion can be extended to systems with multiple VSCs as well as to guide and design VSC control strategy.

## APPENDIX A

The mathematical derivation of the small-signal model of the converter is shown as follows:

**1) The small-signal model of the converter with constant reference values of inner loop**

By linearizing (1)-(3), the small-signal model of the converter can be obtained in the dq frame, which can be shown as follows:

$$\begin{cases} \Delta\theta_{pll} = H_{pll}(s)\Delta U_q = H_{pll}(s)U\Delta\theta_U \\ \Delta\omega = s\Delta\theta_{pll} \end{cases}, \quad (A.1)$$

$$\begin{cases} \Delta U_{sdref} = (\Delta I_{dref} - \Delta I_d)H_i(s) + G_{FF}(s)\Delta U_d - \omega_0 L_f \Delta I_q \\ \Delta U_{sqref} = (\Delta I_{qref} - \Delta I_q)H_i(s) + G_{FF}(s)\Delta U_q + \omega_0 L_f \Delta I_d \end{cases}, \quad (A.2)$$

$$\begin{cases} \Delta U_{sd} = \Delta U_d + L_f s \Delta I_d - \omega_0 L_f \Delta I_q - \Delta\omega L_f I_q \\ \Delta U_{sq} = \Delta U_q + L_f s \Delta I_q + \omega_0 L_f \Delta I_d + \Delta\omega L_f I_d \end{cases}. \quad (A.3)$$

Using coordinate transformation, the currents and the voltages given by dq components can be rewritten in the polar form. For a general phasor $M$, the transformation matrix can be written as:

$$\begin{bmatrix} \Delta M_d \\ \Delta M_q \end{bmatrix} = \begin{bmatrix} \cos\theta & -\sin\theta \\ \sin\theta & \cos\theta \end{bmatrix} \begin{bmatrix} \Delta M \\ M\Delta\theta \end{bmatrix}. \quad (A.4)$$

Therefore, we can rewrite (A.1)-(A.3) in polar form using (A.4) as shown in follows:

$$\begin{cases} \Delta\theta_{pll} = H_{pll}(s)U\Delta\theta_U \\ \Delta\omega = s\Delta\theta_{pll} \end{cases}, \quad (A.5)$$

$$\begin{bmatrix} \Delta U_{sdref} \\ \Delta U_{sqref} \end{bmatrix} = H_i(s)\begin{bmatrix} \Delta I_{dref} \\ \Delta I_{qref} \end{bmatrix} - H_i(s)\begin{bmatrix} \cos\theta_I & -\sin\theta_I \\ \sin\theta_I & \cos\theta_I \end{bmatrix}\begin{bmatrix} \Delta I \\ I\Delta\theta_I \end{bmatrix} + G_{FF}(s)\begin{bmatrix} \Delta U \\ U\Delta\theta_U \end{bmatrix} - \omega_0 L_f \begin{bmatrix} \sin\theta_I & \cos\theta_I \\ -\cos\theta_I & \sin\theta_I \end{bmatrix}\begin{bmatrix} \Delta I \\ I\Delta\theta_I \end{bmatrix}, \quad (A.6)$$

$$\begin{bmatrix} \Delta U_{sd} \\ \Delta U_{sq} \end{bmatrix} = sL_f \begin{bmatrix} \cos\theta_I & -\sin\theta_I \\ \sin\theta_I & \cos\theta_I \end{bmatrix}\begin{bmatrix} \Delta I \\ I\Delta\theta_I \end{bmatrix} + sH_{pll}(s)L_f \begin{bmatrix} 0 & -I_q \\ 0 & I_d \end{bmatrix}\begin{bmatrix} \Delta U \\ U\Delta\theta_U \end{bmatrix} + \begin{bmatrix} \Delta U \\ U\Delta\theta_U \end{bmatrix} - \omega_0 L_f \begin{bmatrix} \sin\theta_I & \cos\theta_I \\ -\cos\theta_I & \sin\theta_I \end{bmatrix}\begin{bmatrix} \Delta I \\ I\Delta\theta_I \end{bmatrix}. \quad (A.7)$$

Considering the aforementioned assumptions, $\Delta U_{sdref} = \Delta U_{sd}$, $\Delta U_{sqref} = \Delta U_{sq}$ is satisfied. Since the reference values for the inner loop are constant, $\Delta I_{dref} = \Delta I_{qref} = 0$ is satisfied. Thus, based on (A.5)-(A.7), the following equation is obtained:

$$0 = sH_{pll}(s)L_f U\Delta\theta_U \begin{bmatrix} -I_q \\ I_d \end{bmatrix} + (1-G_{FF}(s))\begin{bmatrix} \Delta U \\ U\Delta\theta_U \end{bmatrix} + (sL_f + H_i(s))\begin{bmatrix} \cos\theta_I & -\sin\theta_I \\ \sin\theta_I & \cos\theta_I \end{bmatrix}\begin{bmatrix} \Delta I \\ I\Delta\theta_I \end{bmatrix}. \quad (A.8)$$

According to Fig.2, the angles in (A.8) satisfy the following equations:

$$\Delta\theta_U = \Delta\delta - \Delta\theta_{pll}, \quad (A.9)$$
$$\Delta\theta_I = \Delta\varphi - \Delta\theta_{pll}, \quad (A.10)$$

where $\theta_I$ ($\theta_U$) is the phase angle of the current (voltage) in the dq frame; $\varphi$ ($\delta$) is the phase angle of the current (voltage) in the xy frame.

Using (A.9), (A.10), and (A.5), the relationship between $\Delta\theta_{pll}$ and $\Delta\delta$ can be found as:

$$\Delta\theta_{pll} = \frac{H_{pll}(s)U}{1+H_{pll}(s)U}\Delta\delta. \quad (A.11)$$

Using the angle relationships shown in (A.9)-(A.11) and eliminating $\Delta\theta_U$, $\Delta\theta$, and $\Delta\theta_{pll}$ in (A.8), the small-signal model in the dq frame is converted to a small-signal model in the xy frame. The small-signal model of the converter in the xy frame can be written as:

$$\begin{bmatrix} \Delta I \\ I\Delta\varphi \end{bmatrix} = \begin{bmatrix} Y_{g1}(s) & Y_{g2}(s) \\ Y_{g3}(s) & Y_{g4}(s) \end{bmatrix}\begin{bmatrix} \Delta U \\ U\Delta\delta \end{bmatrix}, \quad (A.12)$$

where

$$\begin{cases} Y_{g1}(s) = -\dfrac{1-G_{FF}(s)}{sL_f + H_i(s)}\cos\theta_I \\ Y_{g2}(s) = -\dfrac{1-G_{FF}(s)}{sL_f + H_i(s)}\dfrac{1}{1+H_{pll}(s)U_0}\sin\theta_I \\ Y_{g3}(s) = \dfrac{1-G_{FF}(s)}{sL_f + H_i(s)}\sin\theta_I \\ Y_{g4}(s) = \dfrac{H_i(s)H_{pll}(s)I_0 - (1-G_{FF}(s))\cos\theta_I}{(sL_f + H_i(s))(1+H_{pll}(s)U_0)} \end{cases}.$$

The subscript "0" denotes steady-state values.

**2) The small-signal model of the converter with PQ control**

Linearizing the equation of the PQ controller (8) yields:

$$\begin{bmatrix} \Delta I_{dref} \\ \Delta I_{qref} \end{bmatrix} = H_P(s)\begin{bmatrix} -\Delta P_e \\ \Delta Q_e \end{bmatrix}. \quad (A.13)$$

Based on (A.5)-(A.7) and (A.13), the following equation is obtained:

$$\begin{bmatrix} \Delta P_e \\ \Delta Q_e \end{bmatrix} = \frac{L_f s H_{pll}(s)I}{H_i(s)H_P(s)}\begin{bmatrix} 0 & \sin\theta_I \\ 0 & \cos\theta_I \end{bmatrix}\begin{bmatrix} \Delta U \\ U\Delta\theta_U \end{bmatrix} + \frac{H_i(s)+L_f s}{H_i(s)H_P(s)}\begin{bmatrix} -\cos\theta_I & \sin\theta_I \\ \sin\theta_I & \cos\theta_I \end{bmatrix}\begin{bmatrix} \Delta I \\ I\Delta\theta_I \end{bmatrix}. \quad (A.14)$$

Linearizing (9) and considering the assumption 1, the following equation can be obtained as:

$$\begin{bmatrix}\Delta P_e\\ \Delta Q_e\end{bmatrix}=G_p(s)\begin{bmatrix}U_d & \\ & -U_d\end{bmatrix}\begin{bmatrix}\Delta I_d\\ \Delta I_q\end{bmatrix}+G_p(s)\begin{bmatrix}I_d & \\ & I_d\end{bmatrix}\begin{bmatrix}\Delta U_d\\ \Delta U_q\end{bmatrix}. \quad (A.15)$$

Using the angle relationships in (A.9)-(A.11) and equation (A.15), the small-signal model of the converter in the xy frame can be obtained:

$$\begin{bmatrix}\Delta I\\ I\Delta\varphi\end{bmatrix}=\begin{bmatrix}Y_{g1} & 0\\ 0 & Y_{g4}\end{bmatrix}\begin{bmatrix}\Delta U\\ U\Delta\delta\end{bmatrix}, \quad (A.16)$$

where

$$\begin{cases}Y_{g1}=-\dfrac{H_P(s)H_i(s)G_P(s)I_0}{H_P(s)H_i(s)G_P(s)U_0+H_i(s)+L_f s}\\ Y_{g4}=\dfrac{H_P H_i G_P I_0 - sL_f H_{pll}I_0}{(H_P H_i G_P U_0 + H_i + L_f s)(1+U_0 H_{pll})}+\dfrac{I_0 H_{pll}}{1+U_0 H_{pll}}\end{cases}.$$

**3) The small-signal model of the converter with dc voltage control**

Linearizing the equation of the dc voltage controller (11) yields:

$$\begin{bmatrix}\Delta I_{dref}\\ \Delta I_{qref}\end{bmatrix}=H_{dc}(s)\begin{bmatrix}\Delta U_{dc}\\ 0\end{bmatrix}. \quad (A.17)$$

Linearizing the dynamic equation of the dc capacitor (12) yields:

$$sU_{dc0}\Delta U_{dc}C_{dc}=-\Delta P_e\approx -U_d\Delta I_d - I_d\Delta U_d. \quad (A.18)$$

Using (A.5)-(A.7) and (A.17), and considering the assumption 1, the following equation can be observed:

$$\begin{bmatrix}\Delta P_e\\ 0\end{bmatrix}=\dfrac{s^2 U_{dc0}C_{dc}L_f H_{pll}(s)}{H_{dc}(s)H_i(s)}\begin{bmatrix}0 & 0\\ 0 & -I_d\end{bmatrix}\begin{bmatrix}\Delta U_d\\ \Delta U_q\end{bmatrix}-\dfrac{sU_{dc0}C_{dc}(H_i(s)+sL_f)}{H_{dc}(s)H_i(s)}\begin{bmatrix}1 & 0\\ 0 & 1\end{bmatrix}\begin{bmatrix}\Delta I_d\\ \Delta I_q\end{bmatrix}. \quad (A.19)$$

Using the angle relationships in (A.9)-(A.11) and equation (A.18)-(A.19), the small-signal model of the converter in the xy frame can be obtained:

$$\begin{bmatrix}\Delta I\\ I\Delta\varphi\end{bmatrix}=\begin{bmatrix}Y_{g1} & 0\\ 0 & Y_{g4}\end{bmatrix}\begin{bmatrix}\Delta U\\ U\Delta\delta\end{bmatrix}, \quad (A.20)$$

where

$$\begin{cases}Y_{g1}=-\dfrac{H_{dc}(s)H_i(s)I_0}{sU_{dc0}C_{dc}(H_i(s)+sL_f)+H_{dc}(s)H_i(s)U_0}\\ Y_{g4}=\dfrac{H_i(s)H_{pll}(s)I_0}{(H_i(s)+sL_f)(1+H_{pll}(s)U_0)}\end{cases}.$$

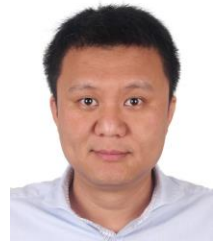

**Huanhai Xin** (M'14) received the Ph.D. degree College of Electrical Engineering, Zhejiang University, Hangzhou, China, in June 2007. He was a post-doctor in the Electrical Engineering and Computer Science Department of the University of Central Florida, Orlando, from June 2009 to July 2010. He is currently a Professor in the Department of Electrical Engineering, Zhejiang University. His research interests include distributed control in active distribution grid and micro-gird, AC/DC power system transient stability analysis and control, and grid-integration of large-scale renewable energy to weak grid.

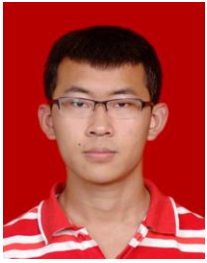
**Ziheng Li** received the B.Eng. degree from Zhejiang University, Hangzhou, China, in 2015. He is currently working toward the M.Eng. degree in the Department of Electrical Engineering, Zhejiang University, Hangzhou, China. His research interests include power system stability and control and grid-integration of large-scale renewable energy to weak grid.

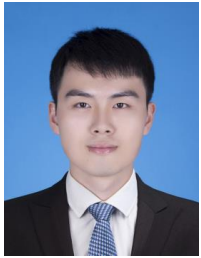
**Wei Dong** received the B.Eng. degree from the Department of Electrical Engineering, Zhejiang University, Hangzhou, China, in 2014, where he is currently working toward the Ph.D. degree. His research interests include power system stability and control and grid-integration of large-scale renewable energy to weak grid.

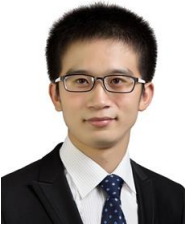
**Leiqi Zhang** received the B.Eng. degree from the Department of Electrical Engineering, Zhejiang University, Hangzhou, China, in 2012, where he is currently working toward the Ph.D. degree. His research interests include distributed generation and its control.

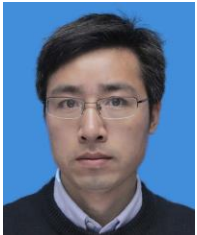
**Zhen Wang** (M'11) received the B.Eng., M.Eng. and Ph.D. degree from Xi'an Jiaotong University, Zhejiang University, and Hong Kong Polytechnic University in 1998, 2001, and 2009, respectively. Currently he is a full-time Professor in the Department of Electrical Engineering, Zhejiang University. He was the recipient of 2014 Endeavour Research Fellowship sponsored by Australia Government and visiting scholar of the University of Western Australia from Feb. 2014 to Aug. 2014. His research interests include power system stability and control, renewable energy integration and VSC-HVDC transmission.

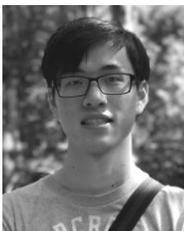
**Jian Zhao** (S'15) received the B.Eng. degree from Zhejiang University, Hangzhou, China, in 2013. He is currently working toward the Ph.D. degree in the Department of Electrical Engineering, The Hong Kong Polytechnic University, Hong Kong. He was a visiting scholar at Argonne National Laboratory, Argonne, IL, USA. He was a recipient of the 2016 IEEE PES Best Paper Reward.
His research interests include distribution network operation and planning, grid integration of electric vehicles and renewable energies.